\documentclass[twoside,11pt]{article}

\usepackage{graphicx}

\usepackage{amssymb}
\usepackage{amsmath}
\usepackage{amsfonts}
\usepackage{wasysym}
\usepackage{revsymb}
\usepackage{amsbsy}
\usepackage{bm}

\usepackage{titlesec}

\usepackage{cite}

\titleformat*{\section}{\bfseries}
\titleformat*{\subsection}{\bfseries}

\begin{document}           

\title{\Large 
Cosmology with the matter component decaying faster than radiation}
\author{V.E. Kuzmichev, V.V. Kuzmichev\\[0.5cm]
\itshape Bogolyubov Institute for Theoretical Physics,\\
\itshape National Academy of Sciences of Ukraine, Kyiv, 03143 Ukraine}

\date{}

\maketitle

\begin{abstract}
The impact of a fast decaying component of mass-energy, that decreases faster than radiation with the increase of the scale factor, on the evolution of the 
universe is studied using a hydrodynamic approach. Proceeding from the Hamiltonian formalism, the hydrodynamic-like equations for the velocity and 
acceleration of the expansion of the universe as a function of conformal time are obtained. The influence of a fast decaying component on the dynamics of the 
expansion of the universe is determined by the sign of its contribution to the total energy density of the system. The effects of this component are illustrated by 
figures. 
\end{abstract}


\section{Introduction}\label{sec:1}
For some time, it has been thought that the limiting case of the equation of state for a perfect fluid or macroscopic bodies treated as continuous is 
$p = \frac{1}{3} \rho$, where $p$ is the pressure and $\rho$ is the energy density \cite{Lan75}. It can be obtained within the framework of special relativity, 
assuming electromagnetic interaction between particles. In cosmological model of the homogeneous and isotropic universe based on general relativity, the 
energy density decreases as $\rho \sim a^{-3 (1 + w)}$, where $a$ is the scale factor, $w = \frac{p}{\rho}$ is the equation of state parameter. Then
for $w = \frac{1}{3}$, the energy density $\rho \sim a^{-4}$ is corresponding to a matter component that decays most rapidly with the 
expansion of the universe. Such a behavior can be attributed to the era of radiation dominance.

The cosmological model in which the very early universe is assumed to be filled with a gas of cold baryons with a stiff equation of state $p = \rho$
has been first introduced in Refs.~\cite{Zel62,Zel72}. When stiff matter dominates, the energy density is $\rho \sim a^{-6}$, so that stiff matter decreases more 
rapidly than radiation.

The same dependence of the energy density $\rho$ on the scale factor emerges within the Einstein--Cartan generalization of the standard cosmological
model, in which the antisymmetric part of 
the affine connection coefficients (torsion) becomes an independent dynamic variable which can be associated with the spin density of matter $s_{\mu \nu}$ in 
the universe. The contribution $\sigma^{2}$ of the macroscopic spin of the fluid to the total energy density is negative and it is given by the square of the spin 
density, $\sigma^{2} = \frac{1}{2} \langle s_{\mu \nu} s^{\mu \nu} \rangle \sim a^{-6}$ \cite{He76,Pon83,Gas86,K13}. In the usual interpretation of the 
Einstein--Cartan theory, the spin density is associated with the quantum-mechanical spin of microscopic particles. The simplest scenario assumes the universe 
is filled with an unpolarized spinning fluid. Elements of quantum description become an integral part of the interpretation of the theory.

Another perspective is possible within the framework of quantum theory of gravity. The quantum corrections to the Einstein--Friedmann equations that follow 
from the application of quantum geometrodynamics for maximally symmetric space can be expressed as the energy density contribution to the total energy 
density in the form $\rho = \frac{Q}{a^{4}}$, where $Q$ is the quantum Bohm potential \cite{K13,K02,K08,K18,K24}. In the model of a single dominant matter 
field we get $Q = \frac{\mu}{a^{2}}$. The numerator $\mu = 2 - \frac{9}{16} (1 + w)(3 - w)$ does not depend on the scale factor $a$ and it determined by the 
equation of state parameter of dominant matter component. Thus the quantum addition to the energy density is equal to $\rho = \frac{\mu}{a^{6}}$.

Introducing such a fast decaying matter (hereafter referred to as FDM) component, that decreases faster than radiation, significantly changes the possible 
scenario of the evolution of the universe. If there was an era of FDM dominance, it should have preceded the eras of dominance of radiation, dust, and dark 
energy. This component would have a significant impact on the conditions for the onset of an inflationary phase of accelerated expansion that can occur either 
at a sufficiently early epoch or later. It should not be disregarded that, under certain conditions, the contribution of the FDM component could be significant not 
only in the early universe \cite{K24}.

In the present note, the impact of an FDM component on the evolution of the universe is studied using a hydrodynamic approach.

\section{Basic equation}\label{sec:2}
The finite degree of freedom model can provide a reasonable framework for addressing cosmological problems.
We will consider the homogeneous and isotropic cosmological system (universe), whose geometry is determined by the Robertson--Walker line element with 
the cosmic scale factor $a$. The Hamiltonian of the model in Planck units is taken as
\begin{equation}\label{1}
H = \frac{N}{2} \left\{- \pi_{a}^{2} - \kappa a^{2} + a^{4} \rho (a) \right\},
\end{equation}
where $N$ is the lapse function which defines the proper time $t$ by the differential equation $dt = a N d \eta$, $\eta$ is the ``arc time'',
$\pi_{a}$ is the momentum canonically conjugate with the variable $a$, the parameter $\kappa = +1, 0, -1$ is the curvature parameter, 
$ \rho (a)$ is the energy density. Equation (\ref{1}) implies averaging over physical fields, spins and other matter degrees of freedom \cite{K13},
so that the energy density has a form
\begin{equation}\label{2}
\rho (a) = \rho_{m} (a) + \rho_{\gamma} (a) + \rho_{\Lambda}  + \rho_{\mu} (a).
\end{equation}
Here $\rho_{m} (a) = \frac{2 M(a)}{a^{3}}$ is the energy density of matter with mass-energy $M(a)$ in a proper volume $\frac{a^{3}}{2}$, which includes 
luminous and dark matter,
$\rho_{\gamma} (a) = \frac{E}{a^{4}}$ is the energy density of radiation (in standard physical units the constant $E$ has the dimensions
$[E] = \mbox{Energy} \times \mbox{Length}$),  $\rho_{\Lambda} = \frac{\Lambda}{3}$ is the contribution from the cosmological constant $\Lambda$,
$\rho_{\mu} (a) = \frac{\mu}{a^{6}}$ is the energy density of the FDM represented by stiff matter, unpolarized spinning fluid or quantum corrections via the 
quantum Bohm potential $Q (a) = a^{4} \rho_{\mu} (a)$. The constant $\mu$ can be either positive (for stiff matter \cite{Zel62,Zel72} and for quantum effects 
under certain conditions \cite{K24}) or negative (for unpolarized spinning field \cite{Pon83,Gas86} and for quantum effects \cite{K13,K24}) when describing 
different physical processes. 

Considering the explicit form of the energy density components, the Hamiltonian can be rewritten as
\begin{equation}\label{3}
H = \frac{N}{2} \left\{- \pi_{a}^{2} - \kappa a^{2} + 2 a M(a) + E + a^{4} \frac{\Lambda}{3} + Q (a) \right\}.
\end{equation}
From the Hamiltonian equations of motion describing how the variables $a$ and $ \pi_{a}$ vary in time $\eta$ we have
\begin{equation}\label{4}
\frac{d a}{d \eta} = \{ a, H \} = - N  \pi_{a},
\end{equation}
\begin{equation}\label{5}
\frac{d  \pi_{a}}{d \eta} = \{ \pi_{a}, H \} = N  \left[\kappa a - M - a \frac{\partial M}{\partial a} - 2 a^{3} \frac{\Lambda}{3} - 
\frac{1}{2} \frac{\partial Q}{\partial a}\right],
\end{equation}
where $\{ . , . \}$ are the Poisson brackets.

The lapse function $N$ plays the role of the Lagrange multiplier which determines the constraint
\begin{equation}\label{6}
- \pi_{a}^{2} - \kappa a^{2} + 2 a M + E + a^{4} \frac{\Lambda}{3} + Q = 0.
\end{equation}
Taking into account Eq.~(\ref{4}) and passing to a time variable $T$, $d T = N d \eta$ (so that $d t = a dT$), we can rewrite Eq.~(\ref{6}) in the form of the 
conservation law
\begin{equation}\label{7}
v^{2} + \kappa a^{2} - a^{4} \rho = 0,
\end{equation}
where $v = \frac{d a}{d T}$ is the velocity at which a cosmological system, considered as a ``particle'' or an infinitesimal volume element, moves
through the minisuperspace. It resembles Bernoulli’s equation for stationary flow of a perfect fluid in a field of conservative forces.

From Eq.~(\ref{5}) follows the equation for the evolution of velocity
\begin{equation}\label{8}
\frac{d v}{d T} = - \kappa a + M + a \frac{d M}{d a} + 2 a^{3} \frac{\Lambda}{3} + \frac{1}{2} \frac{d Q}{d a}.
\end{equation}
This equation can be expressed in a standardized form
\begin{equation}\label{9}
\frac{d v}{d T} = - \kappa a + \frac{a^{3}}{2} \left[\rho - 3 p \right],
\end{equation}
where $\rho$ is the total energy density (\ref{2}), $p = p_{m} + p_{\gamma} + p_{\Lambda} + p_{\mu}$ is the total pressure with the components
\begin{equation}\label{10}
p_{m} = - \frac{2}{3 a^{2}} \frac{d M}{d a}, \quad p_{\gamma} = \frac{E}{3 a^{4}}, \quad p_{\Lambda} = - \frac{\Lambda}{3}, \quad
p_{\mu} = - \frac{1}{3 a^{3}} \frac{d Q}{d a} + \frac{Q}{3 a^{4}}.
\end{equation}

Equation (\ref{9}) can be given a hydrodynamic-like form
\begin{equation}\label{11}
\left(\frac{\partial}{\partial T} + v \frac{\partial}{\partial a} \right) v = \mathcal{F},
\end{equation}
where 
\begin{equation}\label{12}
\mathcal{F} = - \kappa a + \frac{a^{3}}{2} \left[\rho - 3 p \right]
\end{equation}
is the force acting on any given point $a$ of minisuperspace at time $T$, without consideration of which ``particle''  (universe)
occupies the position $a$. This equation can be considered as an analog of Euler's equation for the cosmological system under study.

The law of conservation of energy density in the universe, undergoing expansion in time $T$, can be written as follows:
\begin{equation}\label{13}
\frac{d \rho}{d T} + \frac{v}{a} \left(\rho + p \right) = 0.
\end{equation}
The Hubble expansion rate is expressed in terms of velocity $v$ as $H = \frac{v}{a^{2}}$.

\section{Illustrations}\label{sec:3}
\begin{figure}[!]
\centering
\includegraphics[width=9cm]{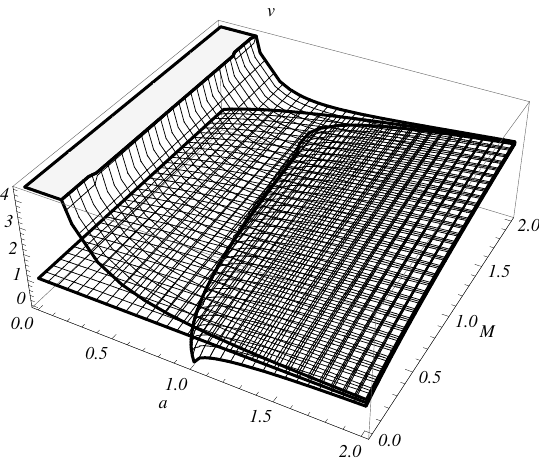}
\caption{The velocity $v$ as a function on $a$ and $M$ at $E =1$ with $\mu = 1$ (upper surface), $\mu = 0$ (middle surface) and $\mu = -1$ (lower surface). 
All units are rescaled by $a_{r}$.}
\label{fig1}
\end{figure}

As an illustration of the influence of the FDM component, which decreases faster than radiation with the increase of the scale factor $a$, we consider the 
simplest case of a spatially flat universe ($\kappa = 0$) without a cosmological constant ($\Lambda = 0$). Fig.~\ref{fig1} shows the evolution of velocity 
$v$ (\ref{7}) as a function on the scale factor $a$ and mass-energy $M$ of dust at fixed values of the parameters $E$ and $\mu$. Each point on the surfaces 
corresponds to one particular universe with specific parameter values. We use the units rescaled to some typical value of the scale factor $a_{r}$ at which the 
contribution of the component $\sim a^{-6}$ is significant. The figure shows that for $\mu > 0$ the contribution of this component at scales $\frac{a}{a_{r}} < 1$ 
becomes prevailing, but its influence extends beyond this region. At the point $a = 0$, the velocity $v$ goes to infinity. In the case $\mu < 0$,  only the real 
values of the velocity $v > 0$ are shown in the plot. Imaginary values of velocity $v$ correspond to imaginary times. This region is described by geometry with 
Euclidean signature \cite{K09,Ha83,HP}.

\begin{figure}[!]
\centering
\includegraphics[width=9cm]{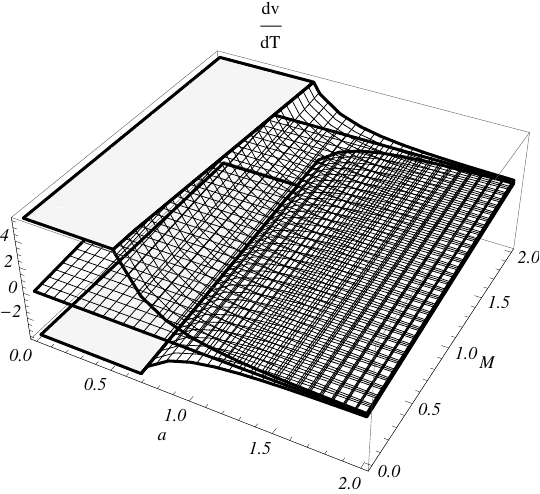}
\caption{The acceleration $\frac{d v}{d T}$ as a function on $a$ and $M$ at $\mu = 1$ (lower surface), $\mu = 0$ (middle surface) and $\mu = -1$ (upper surface). 
All units are rescaled by $a_{r}$.}
\label{fig2}
\end{figure}

More informative from the physical point of view is the behavior of acceleration $\frac{d v}{d T}$ (\ref{8}) as a function of the same parameters, shown in 
Fig.~\ref{fig2}. As this figure shows, in the region $\frac{a}{a_{r}} < 1$, the expansion of the universe is suppressed by the large negative acceleration for 
$\mu > 0$. Beyond this region, the acceleration $\frac{d v}{d T} \sim M$ becomes positive and the universe begins to expand with acceleration. In the case 
$\mu < 0$, the acceleration is positive for all values of $a$, becoming particularly large in the region $\frac{a}{a_{r}} < 1$. The period of accelerated expansion 
can be associated either with the inflationary phase or with the Big Bang itself (at $a_{r} \sim 1$). 

\section{Discussion}\label{sec:4}
To solve the horizon and flatness problems, the cosmological model must include a period of rapid expansion, during which the universe will expand with 
acceleration \cite{Ab84}. The condition of positivity of the second derivative of the scale factor with respect to proper time implies the appearance of a negative 
effective pressure. In models without unpolarized spinning fluid or quantum corrections this can only be achieved with a dominant vacuum contribution to the 
total stress-energy tensor. By introducing an FDM component with  $\mu < 0$, accelerated expansion can be achieved even in the case of 
positive pressure. The possible scenario of the evolution of the universe changes significantly with the inclusion of an FDM component.  As shown in 
Ref.~\cite{K24}, taking an FDM component into account can, in principle, eliminate a discrepancy between the direct late time model-independent 
measurements of the 
Hubble constant and its indirect model dependent estimates known as ``Hubble tension''.

\section*{Acknowledgements}
This work was partially supported by The National Academy of
Sciences of Ukraine (Projects No.~0121U109612 and  No.~0122U000886) and by a grant from the Simons Foundation (USA).

\end{document}